\documentclass[12pt]{article}
\usepackage{amsmath,amssymb,amsfonts}
\usepackage[paper=letterpaper,margin=1.0in]{geometry}
\usepackage{graphicx}
\usepackage{dsfont}

\newcommand{\be}{\begin{equation}}
\newcommand{\ee}{\end{equation}}
\newcommand{\bea}{\begin{eqnarray}}
\newcommand{\eea}{\end{eqnarray}}
\newcommand{\ba}{\begin{array}}
\newcommand{\ea}{\end{array}}
\newcommand{\ben}{\begin{enumerate}}
\newcommand{\een}{\end{enumerate}}
\newcommand{\bi}{\begin{itemize}}
\newcommand{\ei}{\end{itemize}}
\newcommand{\bc}{\begin{center}}
\newcommand{\ec}{\end{center}}
\newcommand{\bfig}{\begin{figure}}
\newcommand{\efig}{\end{figure}}
\newcommand{\bq}{\begin{quotation}}
\newcommand{\eq}{\end{quotation}}
\newcommand{\bt}{\begin{table}}
\newcommand{\et}{\end{table}}
\newcommand{\btab}{\begin{tabular}}
\newcommand{\etab}{\end{tabular}}
\newcommand{\bs}{\begin{slide}}
\newcommand{\es}{\end{slide}}

\begin{document}

{\footnotesize
${}$
}

\bc

\vskip 1.0cm
\centerline{\Large \bf Dark Matter, Infinite Statistics and Quantum Gravity}
\vskip 0.5cm
\vskip 1.0cm

\renewcommand{\thefootnote}{\fnsymbol{footnote}}

\centerline{{\bf
Chiu Man Ho${}^{1}$\footnote{\tt chiuman.ho@vanderbilt.edu},
Djordje Minic${}^{2}$\footnote{\tt dminic@vt.edu}, and
Y.\ Jack Ng${}^{3}$\footnote{\tt yjng@physics.unc.edu}
}}

\vskip 0.5cm

{\it
${}^1$Department of Physics and Astronomy, Vanderbilt University,\\
Nashville, TN 37235, U.S.A.\\
${}$ \\
${}^2$Department
of Physics, Virginia Tech, Blacksburg, VA 24061, U.S.A. \\
${}$ \\
${}^3$Institute of Field Physics, Department of Physics and
Astronomy,\\
University of North Carolina, Chapel Hill, NC 27599, U.S.A.
}

\ec

\vskip 1.0cm

\begin{abstract}

We elaborate on our proposal regarding a connection between global physics
and local galactic dynamics via quantum gravity.
This proposal calls for
the concept of MONDian dark matter which behaves like
cold dark matter at
cluster and cosmological scales but emulates
modified Newtonian dynamics (MOND) at the galactic scale.
In the present paper, we first point out a surprising connection between
the MONDian dark matter
and an effective gravitational Born-Infeld theory. We then argue
that these unconventional quanta of
MONDian dark matter must obey infinite statistics, and the theory must be fundamentally non-local. Finally, we
provide a possible top-down approach
to our proposal from the Matrix theory point of view.

\end{abstract}

\renewcommand{\thefootnote}{\arabic{footnote}}

\vspace{1cm}
~~~PACS Numbers: 95.35.+d; 04.50.Kd; 05.30.-d; 04.60.-m 

\newpage

\section{Introduction}

The fascinating problem of ``missing mass", or dark matter \cite{dark},
has been historically
identified on the level of galaxies. But the
need for dark matter is in fact even more
urgent at larger scales.  Dark matter is apparently required to yield:
(1) the correct cosmic microwave background spectrum shapes (including the
alternating peaks); (2) the correct large-scale
structures; (3) the correct elemental abundances from big bang
nucleosynthesis; and (4) the correct gravitational lensing.
Naturally dark matter has been accorded a prominent place in the concordant
$\Lambda$CDM model of cosmology \cite{dark} according to which cold dark
matter (CDM), dark energy (in the form of
cosmological constant), and ordinary matter account for
about $23 \%$, $73 \%$, and $4 \%$ of the energy and mass of the universe
respectively.

However, at the galactic scale, dark matter does not fare nearly
as well at the larger scales.  It
can explain the observed asymptotic independence of orbital velocities
on the size of the orbit only by fitting data (usually with two
parameters) for individual galaxies.  It is also not very successful in
explaining
the observed baryonic Tully-Fisher relation \cite{TF,McGaugh}, i.e., the
asymptotic-velocity-mass ($v^4 \propto M$) relation.  Another problem
with dark matter is that it seems to possess too much power on small
scales ($\sim 1 - 1000$ kpc) \cite{cen}.

On the other hand, there is
an alternative paradigm that goes by the name of
modified Newtonian dyanmics (MOND) \cite{mond,teves,fmrev},
due to Milgrom.  MOND stipulates that the
acceleration of a test mass $m$ due to the source $M$ is given by
$a= a_N$ for $a \gg a_c$, but
$a = \sqrt{a_N\, a_c}$ for $a \ll a_c$,
where $a_N= G M /r^2$ is the magnitude of
the usual Newtonian acceleration and the critical acceleration
$a_c$ is numerically related to the speed of light $c$ and
the Hubble scale $H$ as
$a_c \approx c H/(2 \pi) \sim 10^{-8} cm/s^2.$
With only one parameter
MOND can explain rather successfully the observed
flat galactic rotation curves and the observed Tully-Fisher relation
\cite{dsmond}.  Unfortunately
there are problems with MOND at the cluster and cosmological scales.

Thus CDM and MOND complement each other well, each being successful
where the other is less so.  We found it natural
to combine their salient successful features
into a unified scheme which straddles the
fields of astronomy and high energy physics.
In our previous work \cite{HMN}, by making use of a novel
quantum gravitational interpretation of (dark) matter's inertia,
we introduced the new concept of MONDian dark matter which behaves like CDM at
cluster and cosmological scales but emulates MOND at the galactic scale.

In this paper, after a short review of our proposal on MONDian dark matter,
we first point out a surprising connection between our proposal and an effective
gravitational Born-Infeld description of the MOND-like phenomenology of
our dark matter quanta.
Furthermore, we stress that these unusual quanta of dark matter must obey the
crucial property of infinite statistics. We illustrate the properties of
an essentially
non-local theory that describes such dark matter with infinite statistics.
We naturally expect that such non-canonical dark matter quanta
should have dramatic signatures
in high energy particle experiments.

\section{From Entropic Gravity to MONDian Dark Matter}

Our previous proposal \cite{HMN}
makes crucial use of a natural
relationship between gravity and thermodynamics
\cite{verlinde,Jacob95}.
The starting point is
the recent work of E. Verlinde \cite{verlinde}
in which the canonical Newton's laws are derived
from the point of view of holography
\cite{hawking,holography,Susskind,adscft}.
Verlinde applies the first law of thermodynamics to propose the concept of
entropic force
\bea
F_{entropic} = T \frac{\Delta S}{\Delta x},
\eea
where $\Delta x$ denotes an infinitesimal spatial displacement of a
particle with mass $m$ from the heat
bath with temperature $T$.
Invoking Bekenstein's original arguments
concerning the entropy $S$ of black holes \cite{bekenstein}
he imposes $
\Delta S = 2\pi k_B \frac{mc}{\hbar} \Delta x
$.\,
With the help of the famous formula for the Unruh temperature,
$
k_B T = \frac{\hbar a}{ 2\pi c},
$\,
associated with a uniformly accelerating (Rindler) observer
\cite{unruh,Davies},
he obtains Newton's second law
$F_{entropic}= T \nabla_x S= m a$.

Next, Verlinde considers an
imaginary quasi-local (spherical) holographic screen of area $A=4 \pi
r^2$ with
temperature $T$. Assuming the equipartition of energy $E=
\frac{1}{2} N k_B T$ with $N$ being
the total number of degrees of freedom (bits) on the screen given by $N =
Ac^3/(G \hbar)$, and employing the Unruh
temperature formula and the fact that $E=M c^2$, he obtains
$
2 \pi k_B T = G M /r^2
$
and recovers exactly the non-relativistic Newton's law
of gravity, namely $a= G M /r^2$.

But we live in an accelerating universe (in
accordance with the $\Lambda$CDM model).
Thus we need a generalization \cite{HMN} of Verlinde's
proposal \cite{verlinde}
to de Sitter space with a positive cosmological constant (which is
related to the Hubble parameter $H$ by $\Lambda \sim 3 H^2$ after setting
$c = 1$).
Since the Unruh-Hawking temperature
as measured by a non-inertial observer with acceleration $a$ in the de
Sitter space is given by
$ \sqrt{a^2+a_0^2}/(2 \pi k_B)$\, \cite{deser,Jacob98}, where
$a_0=\sqrt{\Lambda/3}
$ \,\cite{hawking}, it is natural to
define the net temperature measured by the
non-inertial observer (relative to the inertial observer)
to be
\bea
\tilde{T} = \frac{1}{2\pi k_B} \left(\,\sqrt{a^2+a_0^2}- a_0\,\right)\,.
\eea

In fact, Milgrom has suggested in \cite{interpol} that the difference between the
Unruh temperatures measured by non-inertial and inertial observers in de Sitter space,
namely $2\pi k_B \Delta T =\sqrt{a^2+a_0^2} - a_0$,\,
can give the correct behaviors of the interpolating function between the usual Newtonian
acceleration and his suggested MOND for very small accelerations.
However, he was not able to justify why the force should be related to the difference between the
Unruh temperatures measured by non-inertial and inertial observers in de Sitter space.
Or, in his own words: ``it is not really clear why $\Delta T$ should be a measure of inertia".
As we will see in the following, adopting
Verlinde's entropic force point of view allows us to justify Milgrom's suggestion naturally.

Following Verlinde's approach, the entropic force, acting on the test mass $m$ with
acceleration $a$ in de Sitter space, is obtained by
replacing the $T$ in Verlinde's
argument by $\tilde{T}$ for the Unruh temperature:
\bea
\label{MONDforce}
F_{entropic}=\tilde{T}\, \nabla_x S= m \,\left(\,\sqrt{a^2+a_0^2}-a_0\,\right).
\eea
For $a \gg a_0$, the entropic force is given by $F_{entropic}\approx
ma$, which gives $ a = a_N$ for a test mass $m$ due to the source $M$.
But for $a \ll a_0$, we have
$F_{entropic}\approx ma^2/(2a_0)$; and so
the terminal velocity $v$ of the test mass $m$
should be determined from
\,$ m a^2/(2a_0) = m v^2/r$.

The observed flat
galactic rotation curves (i.e., at large $r$, $v$ is independent
of $r$) and the observed Tully-Fisher relation (the speed of stars
being correlated with the galaxies' brightness,
i.e., $v^4 \propto M$) now require that
$ a \approx \left(\,2 \, a_N \,a_0^3 \,/\pi \right)^{\frac14}$.
\footnote{One can check this
by carrying out a simple dimensional analysis and recalling that
there are two accelerations in the problem: viz, $a_N$ and $a_0$.}
But that means
\bea
F_{entropic} \approx m \frac{a^2}{2\,a_0} = F_{Milgrom} \approx m
\sqrt{a_N a_c}\,,
\eea
for the small acceleration $a \ll a_0$ regime.  Thus
we have recovered MOND
--- provided we identify $a_0 \approx 2 \pi a_c $, with the (observed) critical
galactic acceleration $a_c \sim \sqrt{\Lambda/3} \sim H \sim 10^{-8} cm/s^2$.
Thus, from our perspective, MOND is a phenomenological
consequence of quantum gravity.  To recapitulate,
we have successfully predicted the correct magnitude of the critical galactic
acceleration, and furthermore have found that global physics (in the form of a cosmological constant)
can affect local galactic motion!


Finally, to see how dark
matter can behave like MOND at the galactic scale,
we continue to follow Verlinde's holographic approach
to write $2 \pi k_B \tilde{T} = \frac{G\,\tilde{M}}{r^2}$,
by replacing the $T$ and $M$ in Verlinde's
argument by $\tilde{T}$ and $\tilde{M}$ respectively.
Here $\tilde{M}$ represents the \emph{total} mass enclosed within the
volume $V = 4 \pi r^3 / 3$.
Now it is natural to write the entropic force
$F_{entropic} = m [(a^2+a_0^2)^{1/2}-a_0]$ as
$F_{entropic} = m\,a_N [1 +  (a_0/a)^2/ \pi]$ since the latter expression
is arguably the simplest interpolating formula
for $F_{entropic}$ that satisfies the two requirements: $a \approx (2 a_N
a_0^3/ \pi)^{1/4}$ in the small acceleration $a \ll a_0$
regime, and $a = a_N$ in the $a \gg a_0$ regime.
But we can also write $F$ in another, yet equivalent, form:
$F_{entropic} = G \tilde{M} m/r^2 = G(M+M')m/r^2$,
where $M'$ is some unknown mass --- that is, dark matter.
These two forms of $F$ illustrate the idea of CDM-MOND duality \cite{HMN}.
The first form can be interpreted to mean that there is no dark matter,
but that the law of gravity is modified, while the second form means
that there is dark matter (which, by construction, is consistent with
MOND) but that the law of gravity is not modified.

Dark matter of this kind can behave as if there is no dark matter but
MOND.  Therefore, we call it ``MONDian dark matter"  \cite{HMN}.
Solving for $M'$ as a function of $r$ in the two acceleration
regimes, we obtain $M' \approx 0$ for $a \gg a_0$, and (with $a_0 \sim
\sqrt{\Lambda}$)
\bea
M' \sim (\sqrt{\Lambda}/G)^{1/2}\, M^{1/2}\, r\,,
\eea
for $a \ll a_0$.
This intriguing {\it dark matter profile}
relates, at the galactic scale,
dark matter ($M'$), dark energy ($\Lambda$) and ordinary matter
($M$) to one another. At the moment, it seems prohibitive to check
this prediction in astronomical observations. As a remark, this dark matter profile has been
derived assuming non-relativistic sources and so it is only valid within the galactic scale.
When we enter the cluster or cosmic scale, we need to take into account of the fully
relativistic sources. This may explain why MOND works at the galactic scale, but not at
the cluster or cosmic scale.  One of reasons is that, for the larger
scales, one has to use Einstein's equations with non-negligible
contributions from the pressure and explicitly the cosmological constant,
which have not been taken into account in the MOND scheme \cite{HMN}.

In the above proposal for the dark matter profile, we have assumed spherical symmetry and so it is solely dependent
on $r$. Since both schemes of AQUAL \cite{AQUAL} and QUMOND \cite{QUMOND} reduce to the MOND theory in the spherically symmetric limit,
our proposal should presumably be consistent with AQUAL and QUMOND in that limit. In principle, we could generalize our derivation
to accommodate the general case without spherical symmetry and predict a dark matter disk to compare with AQUAL and QUMOND, but this is certainly
beyond the scope of the present paper.

\section{Gravitational Born-Infeld Theory}

As we have reviewed in the last section, our proposal
combines the MONDian phenomenology with the concept of
dark matter. Since the thermodynamic argument we provided
is highly constrained (as in the formulae for the
effective acceleration and hence the force law), we would like
to use the same constraint to
likewise elucidate the concept of MONDian dark matter.
One way to do this is to look for various reformulations of
MONDian phenomenology.
Given the specific form for the MONDian force law \eqref{MONDforce}, our choices are limited.
One particularly useful reformulation is via an effective
gravitational dielectric medium, motivated by the analogy between
Coulomb's law in a dielectric medium and Milgrom's law for MOND
\cite{fmrev,dielectric}.  As we will show below,
the form of the Born-Infeld Hamiltonian density for electrodynamics
resembles that of the MONDian force law \eqref{MONDforce}. Interestingly, Milgrom
has also noted a similar connection between the nonlinear Born-Infeld electrostatics
and MOND theory \cite{MilgromBornInfeld}. 
Thus the effective gravitational medium for our case is precisely
that of the Born-Infeld type.

Now, we proceed to construct an effective gravitational
Born-Infeld theory and point out its remarkable
connection to the MONDian phenomenology. First of all, the original Born-Infeld (BI) theory \cite{bi}
is defined with the following Lagrangian density (where $\vec{E}$ and $\vec{B}$ are the
respective electric and magnetic fields)
\bea
L_{\rm{BI}}= b^2\, \left(\,1-\sqrt{1- \frac{E^2-B^2}{b^2}-\frac{(\vec{E}\,\cdot\,\vec{B})^2}{b^4}}\,\right)\,,
\eea
where $b$ is a dimensionful parameter. In fact, if we set $\vec{B}=0$, it follows that $b$ represents the maximal field
strength allowed.
The corresponding Hamiltonian density is given by \cite{bi}
\bea
H_{\rm{BI}}= b^2\, \left(\,\sqrt{1+ \frac{D^2+B^2}{b^2}+\frac{(\vec{D}\,\times \,\vec{B})^2}{b^4}}-1\,\right)\,.
\eea
Next, we explore a gravitational analog of the Born-Infeld theory, in which the
relevant field strength is of a gravitational type. In particular, we set $\vec{B}=0$. Then,
the corresponding gravitational Lagrangian and Hamiltonian densities read as
\bea
L_g= b^2\, \left(\,1-\sqrt{1- \frac{E_g^2}{b^2}}\,\right)\,,
\eea
\bea
H_g= b^2\, \left(\,\sqrt{1+ \frac{D_g^2}{b^2}}-1\,\right)\,.
\eea
For our reasoning, the Hamiltonian density is more relevant, and for a
normalization purpose (which will become clear in a moment), we
start from the following normalized Hamiltonian density which has
an extra overall factor of $\frac{1}{4\,\pi}$:
\bea
\label{Hg}
H_g &=& \frac{b^2}{4\,\pi}\, \left(\,\sqrt{1+ \frac{D_g^2}{b^2}}-1\,\right) \nonumber \\
  &=& \frac{1}{4\,\pi}\, \left(\,\sqrt{b^4+ b^2\,D_g^2}-b^2\,\right)\,.
\eea
Let $A_0 \equiv b^2$ and $ A \equiv b \, D_g$, then the Hamiltonian density becomes
\bea
H_g &=& \frac{1}{4\,\pi}\, \left(\,\sqrt{A^2+A_0^2}-A_0\,\right)\,.
\eea
Assuming there exists an energy equipartition, then the effective gravitational
Hamiltonian density, which correspond to the energy, is equal
to
\bea
\label{energy-equipartition}
H_g = \frac{1}{2}\,k_B\, T_{\rm eff}\,,
\eea
where $T_{\rm eff}$ is an effective temperature associated which
the energy through the equipartition of energy.
\footnote{Note that this energy density is energy per
unit volume.  But we can regard it as energy per degree of
freedom by recalling that volume, which usually scales as
entropy $S$, scales as the number of degrees of freedom $N$ in
a holographic setting.  Interestingly $S \sim N$ is one of
the features of infinite statistics \cite{LT}.}
But the Unruh temperature formula implies that
\bea
\label{Teff}
T_{\rm eff} = \frac{\hbar}{2\,\pi\,k_B}\, a_{\rm eff}\,,
\eea
where $a_{\rm eff}$ is the effective acceleration.
As a result, we obtain (after setting $\hbar =1$)
\bea
a_{\rm eff}  = \sqrt{A^2+A_0^2}-A_0\,.
\eea
For a given test mass $m$, the Born-Infeld \emph{inspired} force law
is then given by
\bea
\label{FBI}
F_{\rm BI} = m\, \left(\,\sqrt{A^2+A_0^2}-A_0\,\right)\,.
\eea
Quite remarkably, $F_{\rm BI}$ is of exactly the same form as the
force law \eqref{MONDforce} derived in
our previous paper \cite{HMN} as reviewed in section 2. In what follows,
we give a physical interpretation of this somewhat
formal result and use it to illuminate the properties of the proposed
MONDian dark matter quanta.

\section{MONDian Dark Matter and Infinite Statistics}

In this section, we argue that the surprising connection between
an effective gravitational Born Infeld and the force law \eqref{MONDforce} points
to the concept of infinite statistics
for our MONDian dark matter quanta.
We argue that this is implied by the equivalence principle. Then we
discuss a toy model of a neutral scalar field obeying
infinite statistics as a first step towards a phenomenologically
realistic model of MONDian dark matter.

First, let us use the equivalence principle within the logic of our argument.
In the previous section, the local gravitational fields $\vec{A}$
and $\vec{A}_0$ appeared in the
surprising formal connection between an effective gravitational
Born-Infeld theory
and our MONDian force law \eqref{MONDforce}.
The validity of the equivalence principle suggests that we should identify
(at least locally) the local accelerations $\vec{a}$ and $\vec{a}_0$
with the local gravitational fields
$\vec{A}$ and $\vec{A}_0$ respectively. Namely,
\bea
\vec{a} \equiv \vec{A}, \quad \vec{a}_0 \equiv \vec{A}_0\,.
\eea
In other words, the validity of the equivalence principle suggests that
the temperature $T_{\rm{eff}}$
should be identified as:
\bea
T_{\rm eff} \equiv
\frac{\hbar}{2\,\pi\,k_B}\,\left(\,\sqrt{a^2+a_0^2}-a_0\,\right)\,,
\eea
which, in turn, implies that the Born-Infeld inspired force law takes the form
\bea
\label{FBI-MOND}
F_{\rm BI} = m\, \left(\,\sqrt{a^2+a_0^2}-a_0\,\right)\,,
\eea
which is precisely the MONDian force law derived in \eqref{MONDforce}.
(For consistency, we check that $a_0$, the counterpart of
the constant $b$ in \eqref{Hg}, is itself a constant, being proportional
to $\sqrt{\Lambda}$.)
We thus conclude that
the successful phenomenology of MONDian dark matter may actually be
described in terms of an effective gravitational Born-Infeld theory.
\footnote{We note that, by using the gravitational Born-Infeld
and effective acceleration, we have no need to invoke the gravitational
``bits" in Verlinde's scheme.  Thus, in some sense, we have bypassed
that scheme.}

The gravitational Born-Infeld Hamiltonian $H_g$ is crucially related
to the temperature $T_{\rm{eff}}$ via
the energy equipartition $H_g= \frac{1}{2}\,k_B\, T_{\rm eff}$.
Now, this temperature $T_{\rm eff}$ is obviously very low,
because of the factor of $\hbar$ (see Eq. \eqref{Teff}\,).
As an example, let us consider a typical acceleration of order
10 m$\rm{s}^{-2}$.
The corresponding effective temperature is of order
$T_{\rm eff} \sim 10^{-20}$ K and the effective
characteristic energy scale is of order $k_B\,T_{\rm{eff}} \sim 10^{-24}$ eV.
Obviously, $k_B\,T_{\rm{eff}}$ is much
smaller than even the tiny neutrino masses of order $10^{-3}$ eV or
the mass of any viable
\emph{cold} dark matter candidate which has to be much heavier than 1 eV.

Recall that the equipartition theorem in general states that
the average of the Hamiltonian is given by
\bea
\langle H \rangle = - \frac{\partial \log{Z(\beta)}}{\partial \beta}\,,
\eea
where $\beta^{-1} = k_B T$ and $Z$ denotes the partition function. To obtain
$\langle H \rangle = \frac{1}{2} \,k_B\, T$ per degree of freedom, we require
the partition function to be of the Boltzmann form
\bea
Z = \exp(\,-\beta\, H\,)\,.
\eea
To be a viable cold dark matter candidate, the quanta of
our MONDian dark matter are expected
to be much heavier than $k_B\,T_{\rm{eff}}$.
One may think that it suffices to use the conventional quantum-mechanical
Bose-Einstein or Fermi-Dirac statistics, but
they would not lead to $\langle H \rangle = \frac{1}{2}\, k_B\, T$
per degree of freedom. As a result, the validity
of $H_g= \frac{1}{2}\,k_B\, T_{\rm eff}$ for very low temperature
$T_{\rm{eff}}$ somehow requires a unique
quantum statistics with a Boltzmann partition function.
But this is precisely what is called the infinite
statistics \cite{infinite} as described by the Cuntz algebra
(a curious average of the bosonic and fermionic algebras \cite{infinite})
\bea
a_i \, a^{\dagger}_j = \delta_{ij}\,.
\eea
Thus, by invoking infinite statistics, the assumption of energy equipartition
$H_g= \frac{1}{2}\,k_B\, T_{\rm eff}$, even for
very low temperature $T_{\rm{eff}}$, is justified.

One may reason that the above arguments for infinite statistics
also apply to Verlinde's original proposal \cite{verlinde} which invokes
energy equipartition, and accordingly
he should need introducing infinite statistics
as well. This would be true \emph{if} he assumed that the typical
mass scale of
the quanta of microscopic degrees of freedom (or ``bits" in his
terminology) on the holographic screen is much heavier than $k_B\,T_{\rm{eff}}$.
However, it is \emph{not necessary} for him to make such an assumption,
thereby the requirement for infinite statistics is evaded.
It could well be that the typical mass scale
of the quanta of his ``bits" is much lighter than $k_B\,T_{\rm{eff}}$.
In that case, it is in the high temperature limit, and then he can safely
use the Boltzmann partition function to obtain the energy equipartition
formula. As a result, whether Verlinde requires infinite statistics or not
would not change any of his results.
\footnote{Verlinde only invokes energy equipartition for the ``bits"
(the unknown microscopic degrees
of freedom) on the holographic screen. In his picture, all
matter is emergent from these ``bits" and the emergent particles
can obey infinite statistics or other statistics. But when
ordinary matter particles emerge from
these ``bits", they obey bosonic or fermionic statistics.
How this happens is beyond the scope of this paper.  In short, it
appears that quantum gravitational degrees of freedom obey
infinite statistics though this fact is irrelevant in Verlinde's case.
Nevertheless, we cannot help but wonder whether quantum gravity is
actually the origin of particle statistics and that the underlying
statistics is infinite statistics.  Is it possible that ordinary
particles that obey Bose or Fermi statistics
are actually some sort of collective degrees of freedom?
For a discussion of
constructing bosons and fermions out of particles obeying infinite
statistics, see \cite{bfquon}.}
On the contrary, to be a viable cold dark
matter candidate, the quanta of our MONDian dark matter must be
much heavier than $k_B\,T_{\rm{eff}}$. This means that infinite statistics is
an essential ingredient to our proposal.

Therefore, we have two rather striking observations:\, (i) the relation between
our force law that leads to MONDian phenomenology and an effective
gravitational Born-Infeld theory; (ii) the need for infinite statistics
of some microscopic quanta which underly the thermodynamic description of
gravity implying such a
MONDian force law.

It is natural to ask:
How would infinite statistics mesh with an effective gravitational
Born Infeld theory
and what would such a connection imply for the physical properties of
MONDian dark matter?
Here we recall some facts from string theory as a theory of quantum gravity.
It is well known that in the open string sector, one naturally induces
Born-Infeld theories
\cite{joe}, in general of a non-Abelian kind \cite{nabi}.
Furthermore, in the case of a non-perturbative formulation of
string theory via Matrix theory \cite{bfss} (a light-cone version of
M-theory), it
has been argued that infinite statistics arises naturally \cite{mtinf,LT}.
This Matrix theory is non-Abelian, but is of the Yang-Mills and not
Born-Infeld kind. However, non-Abelian Born-Infeld
like extensions of Matrix theory exist in various backgrounds \cite{nabi},
and thus infinite statistics should naturally emerge
in that context as well.
Thus, from the Matrix theory point of view, we should expect that
infinite statistics and
an effective theory of the gravitational Born-Infeld type are closely related.
This may serve as a top-down justification for the assumption of
the energy equipartition
$H_g= \frac{1}{2}\,k_B\, T_{\rm eff}$ which requires the imposition of
infinite statistics.

As we have argued earlier, with the validity of energy equipartition
and the equivalence principle, the successful phenomenology
of MONDian dark matter could be described in terms of an effective
gravitational Born-Infeld theory which leads to the correct MONDian
force law. But we just showed that the validity of this energy
equipartition requires
some nonstandard degrees of freedom to obey infinite statistics.
It is these nonstandard degrees of
freedom in the effective gravitational Born-Infeld theory that generates the
gravitational fields and leads to the correct MONDian force law.
As is well-known, any modifications to general relativity must either
introduce new local degrees
of freedom or violate the principle of general covariance
(and hence the equivalence principle) \cite{GR}. Since we keep
the equivalence principle
intact in our arguments and do not introduce any new local gravitational
degrees of freedom, we do not modify general relativity. Thus these nonstandard
degrees of freedom in the effective gravitational Born-Infeld theory will
essentially manifest as new particle degrees of freedom.
Since these new particle degrees of freedom when quantized with
infinite statistics leads to the correct MONDian force law, we identify
them as our MONDian dark matter quanta quantized with infinite statistics.

In order to discuss the particle phenomenology of MONDian dark matter, we need
a relativistic field theory of infinite statistics. It is known that any
theory of infinite statistics is
fundamentally non-local \footnote{
That is, the fields associated with infinite statistics are not local,
neither in the sense that their observables commute at spacelike separation
nor in the sense that their observables are pointlike functionals of the fields.}
(albeit consistent with
Lorentz and CPT invariance) \cite{infinite}.
As far as we know, such a complete field-theoretical description of infinite statistics is
not available at present and thus we have to start from scratch.
Here we present a toy model of a neutral scalar field obeying infinite
statistics (see also \cite{previous}) and postpone a full treatment of this
problem to the future.
We start with the Klein-Gordon equation
\bea
\label{eq1}
(\partial^2 + m^2)\,\phi(x) = 0\,.
\eea
Since $\phi$ is a Hermitian field operator, it can be expanded as
\bea
\label{eq2}
\phi(x) = \int d \omega_k \left(\, a(\vec{k})\, e^{-i k \cdot x} +
 a^{\dagger} (\vec{k})\, e^{i k \cdot x}\,\right)\,,
\eea
where $d \omega_k \equiv \frac{d^3 k}{(2 \pi)^3 2 \sqrt{\vec{k}^2
+ m^2}}$ with $k \cdot x \equiv  \sqrt{\vec{k}^2
+ m^2}\, t - \vec{k} \cdot \vec{r}$.  The annihilation operator $a$ and
creation operator $a^{\dagger}$ obey the infinite statistics algebra
\bea
\label{eq3}
a (\vec{k}) a^{\dagger} (\vec{k'}) = 2\, k^0 \,(2\, \pi)^3\, \delta^{(3)} ( \vec{k}
- \vec{k'})\,,
\eea
where $k^0 \equiv \sqrt{ \vec{k}^2 + m^2}$, and
\bea
\label{eq4}
a(\vec{k}) \,| 0 \rangle = 0 = \langle 0 |\, a^{\dagger} (\vec{k})\,.
\eea
The Wightman function is given by
\bea
\label{eq5}
\Delta^{(+)}(x-y) &=& \langle 0 | \phi (x) \phi (y) | 0 \rangle \nonumber \\
                &=& \int d \omega_k\, e^{-i k \cdot (x - y)},
\eea
where we have used Eqs. (\ref{eq2} - \ref{eq4}).  The Feynman propagator
\bea
\label{eq6}
\Delta_F (x-y) \equiv  -i\,\langle 0 | T (\phi(x) \phi(y) ) | 0 \rangle\,,
\eea
is given, in terms of the Wightman functions, by
\bea
\label{eq7}
\Delta_F (x-y) &=& -i\,\theta (x^0 - y^0) \,\Delta^{(+)} (x-y) -i\, \theta (y^0 -
   x^0)\, \Delta^{(+)} (y-x) \\
   \label{eq8}
               &=& \int \frac{d^4 k}{(2 \pi)^4} \frac{ e^{- i k \cdot
   (x - y)}} {k^2 - m^2 + i\epsilon}\,.
\eea
Eq. (\ref{eq8}) can be shown to be equal to Eq. (\ref{eq7}) by two
different ways.  One way is
by explicitly performing the integration over $k_0$ in Eq.(\ref{eq8}) to yield
Eq. (\ref{eq7}).  Another way is to show that the Feynman propagator is a Green's
function for the Klein-Gordon equation, i.e.,
\bea
(\partial_x^2 + m^2) \,\Delta_F(x-y) = - \delta^4 (x-y),
\eea
by applying $(\partial_x^2 + m^2)$ on Eq. (\ref{eq7}) and using the fact that
the Wightman function solves the Klein-Gordon equation and that, with
the aid of Eq. (\ref{eq5}),
$\Delta^{(+)} (x-y) = \Delta^{(+)} (y-x)$ at equal time $x^0 = y^0$.

From Eq. (\ref{eq8}), it is obvious that we get back the conventional result
for the scalar propagator and non-locality is not
manifest. Mathematically, this is because only the term
$ \langle 0| a \,a^\dagger |0 \rangle $ gives a non-zero
contribution to the propagator. The commutator of $a$ and $a^\dagger$,
which is absent in the infinite statistics case, is not
required for the calculation of the propagator.
But while non-locality
is not manifest in the propagator, it is not completely lost.
The reason is that the equal-time commutator
$[\,\phi(x), \,\phi(y)]|_{x^0=y^0}$ is non-zero, which is a manifestation of
non-locality.

We thus conclude that this toy model is illuminating to some extent,
and indeed it could serve as a preliminary model
for MONDian dark matter. However, we will not explore it further.
This is because we are more interested
in the particle physics phenomenology of the non-locality associated
with MONDian dark matter. Such phenomenological studies will crucially rely
on a non-local propagator of the infinite statistics quanta, as well as
their interactions with the Standard Model particles. In contrast,
non-locality in this toy model is only manifest in the equal-time
commutator $[\,\phi(x), \,\phi(y)]|_{x^0=y^0}$; but it is not
clear that it will
lead to any direct and observable phenomenological consequences. Undoubtedly,
for the particle physics phenomenology of MONDian dark matter to be
relevant, we will need a full description which involves
a truly non-local field theory of infinite statistics quanta.
Investigating the precise nature of such a non-local theory is the next
step in our research program. However, the proposal that MONDian
dark matter quanta should be described by
a non-local theory of infinite statistics, with ultimate origins in
quantum gravity, is already quite remarkable, and this feature of MONDian dark
matter uniquely distinguishes our suggestion from other phenomenological
models of dark matter.

We end this section with the following observation on the phenomenology of
MONDian dark matter. On the one hand, infinite statistics has been associated with
the physics of quantum gravitational quanta such as D0-branes in
particular backgrounds \cite{mtinf}
as well as with black hole physics (as in the work of Strominger \cite{mtinf}).
On the other hand, there are
existing proposals arguing for the relevance
of primordial black holes in the physics of dark matter \cite{primbh},
and, what is more important, for experimental searches for such primordial black holes
\cite{primbh}.  Naturally we are led to conjecture that the
application of the same experimental techniques may be relevant in the observational search of
MONDian dark matter with infinite statistics.

\section{Conclusion:\, Infinite Statistics and Quantum Gravity}

In this paper, we have further developed our proposal for MONDian dark matter which
unifies the salient features of cold dark matter (CDM)
and the phenomenology of
modified Newtonian dynamics (MOND). The MONDian dark matter behaves like CDM at
cluster and cosmological scales but emulates MOND at the galactic scale.
We have pointed out a surprising connection between our
proposal and an effective
gravitational Born-Infeld description of the MOND-like phenomenology of
our dark matter quanta.  Furthermore,
we have argued that these unusual quanta of dark matter must obey the
crucial property of infinite statistics. Thus, MONDian dark matter has to be
described as an essentially
non-local theory for such infinite statistics quanta.
Such a theory would be fundamentally quantum gravitational and thus
distinguished from the usual phenomenological models of dark matter.

We conclude by presenting a possible top-down approach
to our proposal. As already mentioned, it is quite natural to expect that quantum gravity in some form
of Matrix Theory \cite{bfss}, has a non-abelian Born Infeld extension.
If one concentrates on the $U(1)$ part of that theory, which would
correspond to a ``center of mass'' sector of the full quantum theory of gravity,
one will in principle expect to derive an effective gravitational Born-Infeld theory of the kind
discussed in this paper.
Also Matrix theory \cite{mtinf} allows for infinite statistics being a
theory of large (infinite size) matrices.
Thus it would be possible to envision a gravitational Born-Infeld
Hamiltonian which, in conjuction with the equipartition theorem
that is true for infinite statistics, would imply the temperature
formula and thus the force law derived in our previous paper \cite{HMN}.
Finally, by invoking the equivalence principle in this thermodynamic
limit, we would be able to derive the exact formula
$ [\sqrt{a^2+a_0^2}-a_0]$ from which we could deduce the
MONDian scaling at galactic distances.

This scenario would imply that quantum gravity (in the guise of
M-theory) is really behind
MONDian dark matter and its implications for particle physics as
well as astronomy on the galactic and extragalactic scales.
In this discussion, we would need to take account of
holography (i.e. a matrix model description)
in the cosmological asymptotically de Sitter background \cite{vijay}, which will
be quite non-trivial.
One simple idea would be to envision a matrix model (inspired by Matrix theory \cite{bfss})
\bea
L = Tr \left( \,\frac{1}{2}\,(\,\partial\, \mathds{M}\,)^2 + m^2\, \mathds{M}^2 + g \,V(\mathds{M})\, O_{SM} \,\right)\,,
\eea
where $\mathds{M}$ is an infinite dimensional square matrix. The mass term $m^2$
and the ``Yukawa'' coupling $g$ are phenomenological parameters.
Here $V(\mathds{M})$ denotes some effective potential (for simplicity, we can envision a
quartic term $\mathds{M}^4$) and $O_{SM}$ is the relevant standard model
operator that describes the necessary coupling to the dark matter sector.
The mass parameter $m$ could be related to the cosmological SUSY breaking
mechanism of Banks \cite{banks} if the matrix $\mathds{M}$ has
fundamental origins in Matrix theory
in a cosmological de Sitter background. However this topic is beyond
the scope of our
present work and we leave it for further study in the future.

\vskip 0.5cm

\noindent
{\bf Acknowledgments:}
We thank many colleagues
for comments on our proposal concerning MONDian dark matter \cite{HMN}.
In particular we acknowledge useful discussions with and communications from
N. Arav,
V. Balasubramanian, A. Berlind, L. N. Chang, D. Hooper, P. Huber, T. Kephart,
J. Khoury,
A. Kosowsky, D.S. Lee, C.N. Leung,
D. Marfatia,  M. Milgrom, S. Pakvasa, D. Pavon, L. Piilonen, J. Simonetti,
T. Takeuchi, M. Trodden, E. Verlinde,
T. Weiler, E. Witten and J. Zupan.
CMH, DM, and YJN are supported in part by the US Department of Energy
under contract DE-FG05-85ER40226,
DE-FG05-92ER40677 and DE-FG02-06ER41418 respectively.
Finally, we also thank an anonymous referee for calling our attention to Ref. \cite{MilgromBornInfeld}.


\begin{thebibliography}{25}

\bibitem{dark}
For a recent review, see G. Bertone, D. Hooper and J. Silk,
Phys. Rept. {\bf 405}, 279 (2005), and references therein.

\bibitem{TF}
R.B. Tully and J.R. Fisher, Astronomy and Astrophysics {\bf 54},
661 (1977).

\bibitem{McGaugh}
S.S. McGaugh et al., Astrophysical Journal {\bf 533}, L99 (2000).

\bibitem{cen}
R. Cen, arXiv:astro-ph/0005206.

\bibitem{mond}
M. Milgrom, Astrophys. J. {\bf 270}, 365, 371, 384 (1983).

\bibitem{teves}
J. D. Bekenstein, Phys. Rev. {\bf D70}, 083509 (2004).

\bibitem{fmrev}
For an exhaustive review of MOND, see B. Famaey and S. McGaugh,
arXiv:1112.3960.

\bibitem{dsmond}
M. Milgrom, Astrophys. J. {\bf 698}, 1630 (2009).

\bibitem{HMN}
C. M. Ho, D. Minic and Y.J. Ng, Phys. Lett. {\bf B693}, 567 (2010)
[arXiv:1005.3537 [hep-th]];
Gen.\ Rel.\ Grav.\  {\bf 43}, 2567 (2011) [arXiv:1105.2916 [gr-qc]].

\bibitem{verlinde}
E. Verlinde, JHEP {\bf 1104}, 029 (2011) [arXiv:1001.0785 [hep-th]].

\bibitem{Jacob95}
T. Jacobson, Phys. Rev. Lett. {\bf 75}, 1260 (1995).  Also see
T.~Padmanabhan,
  Mod.\ Phys.\ Lett.\ A {\bf 25}, 1129 (2010)
  [arXiv:0912.3165 [gr-qc]]; L. Smolin, arXiv:1001.3668.

\bibitem{hawking}
S.W. Hawking, Comm. Math. Phys. {\bf 43}, 199 (1975).

\bibitem{holography}
G. 't Hooft, arXiv: gr-qc/9310026.

\bibitem{Susskind}
L. Susskind, J. Math. Phys. {\bf 36}, 6377 (1995).

\bibitem{adscft}
O. Aharony, S.S. Gubser, J. Maldacena, H. Ooguri, Y. Oz,
Phys. Rept. {\bf 323}, 183 (2000), and references therein.

\bibitem{bekenstein}
J. D. Bekenstein, Phys. Rev. {\bf D7}, 2333 (1973).

\bibitem{unruh}
W. G. Unruh, Phys. Rev. {\bf D14}, 870 (1976).

\bibitem{Davies}
P.C.W. Davies, J. Phys. {\bf A8}, 609 (1975).

\bibitem{deser}
S. Deser and O. Levin, Class. Quant. Grav. {\bf 14}, L163 (1997).

\bibitem{Jacob98}
T. Jacobson, Class. Quant. Grav. {\bf 15}, 251 (1998).

\bibitem{interpol}
M. Milgrom, Phys. Lett. {\bf A253}, 273 (1999).

\bibitem{AQUAL}
  J.~Bekenstein and M.~Milgrom,
  Astrophys.\ J.\  {\bf 286}, 7 (1984).

\bibitem{QUMOND} M.~Milgrom,
  Mon.\ Not.\ Roy.\ Astron.\ Soc.\  {\bf 403}, 886 (2010)
  [arXiv:0911.5464 [astro-ph.CO]].

\bibitem{dielectric}
L. Blanchet, arXiv:astro-ph/0605637.

\bibitem{MilgromBornInfeld}
M. Milgrom, J. Phys. A {\bf 35}, 1437 (2002) [arXiv:math-ph/0112040].

\bibitem{bi}
For a modern review of Born-Infeld theory, including its relation to
string/M-theory, see
G. W. Gibbons, Rev. Mex. Fis. {\bf 49} S1, 19 (2003) and references therein.


\bibitem{LT}
H. Liu and A.A. Tseytlin, JHEP {\bf 980}, 010 (1998).

\bibitem{infinite}
S. Doplicher, R. Haag and J. Roberts, Commun. Math. Phys. {\bf 23}, 199
(1971);
{\bf 35}, 49 (1974);
A.B. Govorkov, Theor. Math. Phys. {\bf 54}, 234 (1983);
O. Greenberg, Phys. Rev. Lett. {\bf 64}, 705 (1990);
also arXiv:cond-mat/9301002.

\bibitem{bfquon}
O.W. Greenberg and J.D. Delgado, arXiv:hep-th/0107058.

\bibitem{joe}
For an insightful review of string theory see, for example:
J. Polchinski, {\it String Theory}, Cambridge University Press, 1998.

\bibitem{nabi}
On the Non-Abelian Born-Infeld theory in string theory, consult:
A. A. Tseytlin, Nucl. Phys. {\bf B 501}, 41 (1997); R. C. Myers,
JHEP {\bf 9912}, 022 (1999);
R. C. Myers, Class. Quantum Grav. {\bf 20}, S347 (2003)
and references therein.

\bibitem{bfss}
T. Banks, W. Fischler, S.H. Shenker and L. Susskind,
Phys. Rev. {\bf D 55}, 5112 (1997).

\bibitem{mtinf}
T. Banks, W. Fischler, I.R. Klebanov and L. Susskind,
Phys. Rev. Lett. {\bf 80}, 226 (1998); JHEP {\bf 9801}, 008 (1998);
D. Minic, arXiv:hep-th/9712202.
Also, on quantum gravity and infinite statistics, see:
A. Strominger, Phys. Rev. Lett. {\bf 71}, 3397 (1993); I.V.Volovich,
arXiv:hep-th/9608137.

\bibitem{GR}
K.~Kuchar,
  J.\ Math.\ Phys.\  {\bf 15}, 708 (1974);
S.~A.~Hojman, K.~Kuchar and C.~Teitelboim,
  Annals Phys.\  {\bf 96}, 88 (1976).

\bibitem{previous}
Previous work on toy models of infinite statistics and dark energy/dark matter:
Y.~J.~Ng, Phys. Lett. {\bf B 657}, 10 (2007) [arXiv:gr-qc/0703096];
V. Jejjala, M. Kavic and D. Minic,
Adv. High Energy Phys. {\bf 2007}, 21586 (2007).

\bibitem{primbh}
See, e.g., P.H. Frampton, M. Kawasaki, F. Takahashi and T. Yanagida, arXiv:1001.2308[hep-ph];
D.B. Cline et. al., arXiv:0704.2398[astro-ph]; D.B. Cline et al., arXiv:1105.5363.


\bibitem{vijay}
  V.~Balasubramanian, J.~de Boer and D.~Minic,
  Class.\ Quant.\ Grav.\  {\bf 19}, 5655 (2002)
  [Annals Phys.\  {\bf 303}, 59 (2003)].
See also, arXiv:gr-qc/0211003 as well as
D. Minic and C. H. Tze, Phys. Rev. {\bf D 68}, 061501 (2003); Phys. Lett. {\bf B 581}, 111 (2004).

\bibitem{banks}
T. Banks, 2010 TASI lectures, arXiv:1007.4001v3 [hep-th].


\end{thebibliography}
\end{document}